\documentclass[10pt,comsoc,letterpaper]{IEEEtran}

\usepackage[utf8]{inputenc}
\usepackage{amsmath}
\usepackage{mathtools}
\usepackage{amsthm}
\usepackage{times}
\usepackage{algorithm}
\usepackage[noend]{algorithmic}
\usepackage{algorithmic}
\usepackage{graphicx}
\usepackage{verbatim}
\usepackage{amssymb}
\usepackage{cite}
\usepackage{epsfig}
\usepackage{color}
\usepackage{url}
\usepackage{xcolor,colortbl}
\usepackage{multirow}
\usepackage{grffile}
\usepackage{epstopdf}
\usepackage{soul}

\usepackage{listings}
\usepackage[T1]{fontenc}

\begin{document}

\title{A High-Level Rule-based Language for Software Defined Network Programming based on OpenFlow}

\author{

\IEEEauthorblockN{Mehdi Mohammadi, Ala Al-Fuqaha, Zijiang James Yang}

\IEEEauthorblockA{\textit{Computer Science Department} \\
\textit{Western Michigan University}\\
\{mehdi.mohammadi, ala.al-fuqaha, zijiang.yang\}@wmich.edu}
}

\IEEEpubid{\begin{minipage}{\textwidth}\ \\[30pt] \centering
  This paper has been presented in the poster section of GENI Engineering Conference 22 (GEC 22), Washington D.C., March 23-26, 2015.
\end{minipage}} 

\IEEEcompsoctitleabstractindextext{
\begin{abstract}
This paper proposes XML-Defined Network policies (XDNP), a new high-level language based on XML notation, to describe network control rules in Software Defined Network environments. We rely on existing OpenFlow controllers specifically Floodlight but the novelty of this project is to separate complicated language- and framework-specific APIs from policy descriptions. This separation makes it possible to extend the current work as a northbound higher level abstraction that can support a wide range of controllers who are based on different programming languages. By this approach, we believe that network administrators can develop and deploy network control policies easier and faster.
\end{abstract}

}

\maketitle

\IEEEdisplaynotcompsoctitleabstractindextext
\IEEEpeerreviewmaketitle

\section{Introduction}\label{sec:Introduction}
By Software-Defined Network (SDN) technology, network engineers and administrators can control and manage network services through abstraction of lower level functionality. This end can be achieved by splitting the system that makes decisions where to send the packets (control plane) from the underlying systems known as data plane that is in charge of forwarding packets to the selected destinations. In order to be practical, SDN needs some mechanism for the control plane to interact with the data plane. OpenFlow \cite{mckeown2008openflow} is such a protocol that has been used in network research community in recent years. There are couples of OpenFlow controller frameworks such as POX\footnote{http://www.noxrepo.org/pox/about-pox/}, NOX\footnote{http://www.noxrepo.org/}, Beacon\footnote{https://openflow.stanford.edu/display/Beacon/Home}, Floodlight\footnote{http://www.projectfloodlight.org/}, Trema\footnote{http://trema.github.io/trema/}, NodeFlow\footnote{http://garyberger.net/?p=537} and Ryu\footnote{http://osrg.github.io/ryu/}. 

The emerge of OpenFlow simplified network management by providing high-level abstractions to control a set of switches remotely. An OpenFlow framework requires network engineers or administrators to write programs to control and manage data traffic in their network and control network devices. However, one potential problem that network engineers face is that the programming languages that support OpenFlow are complex and administrators are required to know much irrelevant information to develop and deploy a control policy.  The problem will be more prohibitive for a beginner network engineer who does not have a good background in the programming language of the controller. There are other challenges for programmers including \cite{foster2011frenetic}: the interaction between concurrent modules, low-level interface to switch hardware, and multi-tiered programming model. A simple and unified language in a higher abstraction level that does not depend on a specific language can fill this gap. We have developed a text-based format based on XML by which a human-friendly semantic of operations is developed for policy description. XML has been used widely in network management and configuration protocols like NETCONF. This work can be annexed to the current OpenFlow controller as a top layer service. 

Two contributions of this paper are: 
\begin{itemize}
\item proposing an XML-based script language for describing network control policies; and
\item implementing a translator that converts the XML file to a Java source code containing the controller program for Floodlight.
\end{itemize}
The emulation experiments affirms the applicability of this XML notation as a policy describer in SDNs.

\section{Related Work}\label{sec:related_work}
XML documents are widely used to describe systems, to configure or control them. One attractive usage is code generation. There are several works that try to use XML notation as a representation of source codes. JavaML \cite{badros2000javaml} is such a work that represents Java source code in XML notation. The source code representation in JavaML is in a way that constructs like superclasses, methods, message sends, and literal numbers are all directly represented in the elements and attributes of the document content. XML notation is also used in another work named srcML \cite{maletic2002source} by which structural information is added to unstructured source code files. srcML aims to enhance source code representation by adding syntactic information obtained from parse tree. 

In \cite{liang2012nedl}, Liang \textit{et al.} proposed a proof of concept to use XML as a description scheme for OpenFlow Networking experiments. They defined networking experiments in a hierarchical model in which the experiment is at the highest level, and each experiment contains information, topology, deployment, control, and output components. However, they just provided the format description, but not mentioned by which way they generated their XML parser. Furthermore, they have not provided a full evaluation of their system in the real environments. Our work differs from their work in the way we design a platform by which network engineers can describe their network control rules by XML notations regardless of the underlying topology or network elements. In other words, Liang’s work focuses on description of network environment not the control of network traffic and behavior.

There are several works and projects aim to propose a higher level abstraction above the OpenFlow APIs in their development frameworks \cite{foster2011frenetic,monsanto2012compiler,monsanto2013composing,anderson2014netkat}. Frenetic \cite{foster2011frenetic} which has been implemented in python emerged with some simple rules including predicate-action pairs, in which actions support filtering, forwarding, duplicating, and modifying packets. Later it included other more complicated operators like packet processing functionalities \cite{monsanto2012compiler}. Pyretic \cite{monsanto2013composing} as a modern SDN programming language based on Frenetic and beyond the current parallel composition operator, presents two more complex abstractions: sequential composition operator and applying control policies over abstract topologies. By these abstractions the development of modular control programs becomes simpler. NetKat \cite{anderson2014netkat} is another language for SDN with a solid mathematical foundation that supports primitives like filtering, modifying, and transmitting packets as well as union and sequential composition operators. 

There are other works that have more concentration to provide a high-level language for policy description. Procera \cite{voellmy2012procera} and FML \cite{hinrichs2009practical} are designed based on this aim. The declarative policy language in Procera is based on functional reactive programming. This domain-specific language relies on Haskell and is able to use the constructs and data types defined in Haskell. FML has been built on Datalog language to support a declarative logic-based programming interface. In \cite{smith2014management}, as part of their resilient network management system, authors have proposed a policy description language based on management patterns. The management patterns describe the handling and controlling of individual OpenFlow resilience services.

\section{Software Defined Networking}\label{sec:sdn}

Software-Define Networking defines two separate layers as the new architecture of network environments. A data plane that is supposed to do operations like buffering packets, forwarding, dropping, tagging and collecting packet statistics. Control plane, on the other hand, may have the algorithms to track the dynamic topology of the network and has route processing capability. The control plane usually consists of a separate powerful machine called controller. The control plane uses its computed data and the data plane’s statistics to manage and govern a set of dependent switches by installing or removing packet forwarding rules over them. 

OpenFlow as a realization of SDN follows this two-layer architecture. In a typical OpenFlow network, if a switch can find a rule match to the received packet in its flow table, then it proceeds with that rule. Otherwise, the packet is sent to the controller for more processing. The controller examines packet header and establishes a rule based on that packet. The next similar packets arrived in the switches then are not required to go to the controller since the switches have appropriate rule to process them. Although forwarding packets to the controller increases their latency but it occurs not very much.

\section{The Proposed System}\label{sec:system}
\begin{figure}[t]	
	\begin{center}		
		\includegraphics[width=.45\textwidth]{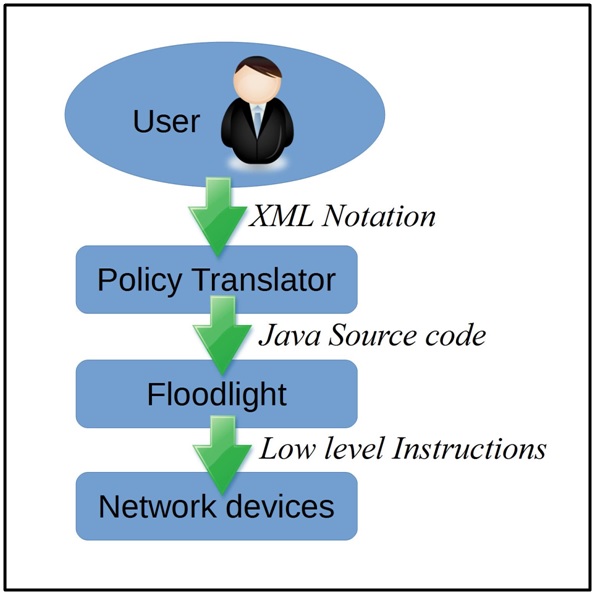}		
	\end{center}	
	\caption{The overall architecture of the system.}\label{fig:bigpic}	
\end{figure}

Our XML translator consists of a lexical analyzer (lexer) and a syntax analyzer (parser). The lexer tokenizes the input file and matches the tags, attributes, identifiers, constant values and so on based on regular grammars. Syntax analyzer, on the other hand, performs syntactic analysis of the input file and if it does not find any problem in this step, generates appropriate Java source code. The overall architecture of the system is depicted in Figure~\ref{fig:bigpic}.

\subsection{Language Specification}
\begin{figure}[t]	
	\begin{center}		
		\includegraphics[width=.45\textwidth]{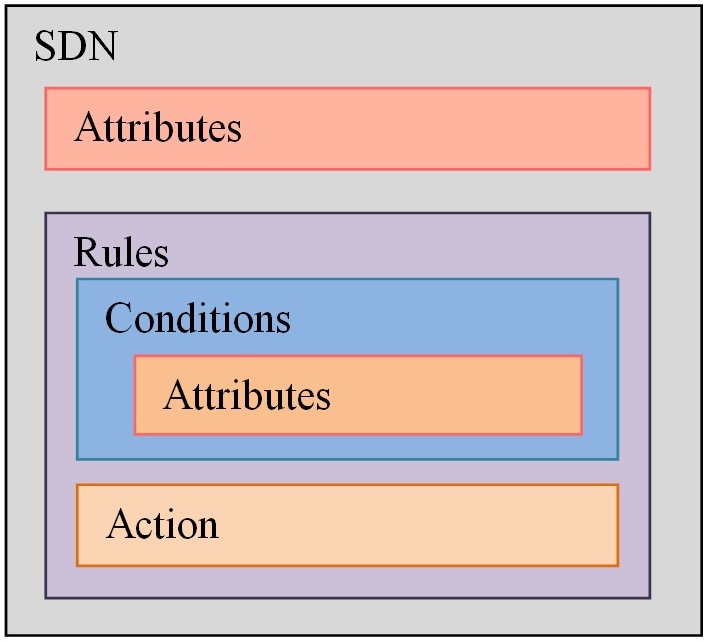}		
	\end{center}	
	\caption{The format of a policy description file.}\label{fig:policy}	
\end{figure}

\begin{figure}[t]	
	\begin{center}		
		\includegraphics[width=.45\textwidth]{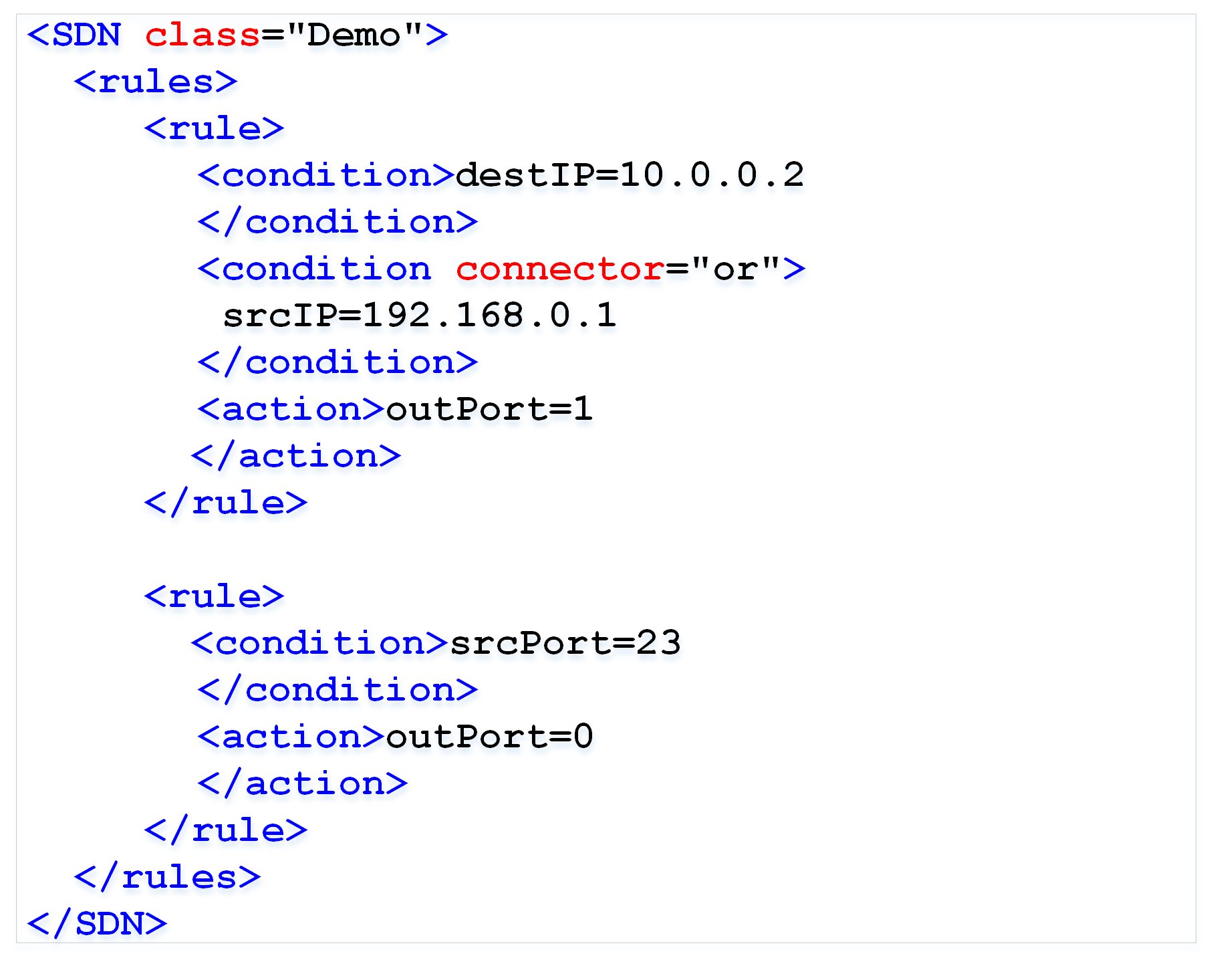}		
	\end{center}	
	\caption{A sample XML policy description.}\label{fig:code}	
\end{figure}

The overall design of an XML file to be used as control program should follow the format of figure~\ref{fig:policy}. Rule description is defined in a hierarchical structure in which at the top level, we define the class name in the SDN element. Then a list of rules contacting zero or more rule elements should be declared. Inside each rule, one or more conditions are expressed. To have compositional conditions, condition elements support logical operators which are stated by attribute “connector”. For example, if a condition element has an “or” connector, it will be joined to the previous condition by logical “or” operator in java source code. The conditions themselves comply with a simple pattern “variable op value”. Variables can be chosen from \texttt{src{\_}ip} (source IP), \texttt{dest{\_}ip} (destination IP), \texttt{src{\_}prt} (source port) and \texttt{dest{\_}prt} (destination port). The current supported operator is equal sign. Value can be a port number or IP address. An example of XML file is shown in Figure~\ref{fig:code}.

The XML sample contains two rules. The first one says that all the packets who are going to IP address 10.0.0.2 or who they are coming from IP address 192.168.0.1 should be forwarded to port 1 of the switches (which assigned to IP address 10.0.0.1). The second rule indicates that each packet originated from Telnet service (port 23) should be dropped (specified by port 0). This example shows that it is possible to implement all network policy management schemes and services like firewalls or load balancing with this XML notation.

\subsection{Lexical Analyzer}
We designed our lexer with LEX format by which we defined the matching patterns for XML elements, attributes and literals that are needed in a typical network controller. The source of lexer file has to be fed to Flex to generate a source code in C language. 
We used regular expression notations to construct the lexer. For example, to define the pattern for IP address, we define these regular expressions in the lexer file:

oct10 	[1-9][0-9]|1[0-9][0-9]|2[0-4][0-9]|25[0-5] 

start{\_}oct 	[1-9]|{oct10}

octet 	[0-9]|{oct10}

ipv4 		{start{\_}oct}\.{octet}\.{octet}\.{octet}

Tag names are considered as keywords in the script and are required to comply exactly with the XML specification.

\subsection{Syntax Analyzer}
The grammar section of translator is defined in syntax analyzer. We used YACC tool to generate the translator. In its input file, we define all the tokens that are introduced in the lexer, grammar production rules that examine the syntax of the XML file and the output associated with each production rule leading to code generation for the XML file. The syntax analyzer description follows the Backus–Naur Form (BNF) notation for context-free grammars. For example, the following production rules show the nesting and iteration nature of XML script for control policies:

\begin{lstlisting}

ruleslist:
	ruleslist rule 
	|rule			
	| /* nothing */
	;
rule:
	RULE_S {fprintf(fp, "\t\tif "); 
    conditionCounter = 0;} 
    conditionList action RULE_E
	;
		
conditionList:
	conditionList condition
	| condition
	;
condition:
	COND_S GT { conditionCounter++; } 
    condText COND_E		
	| ...
condText:		
		SD_IP EQ IPv4 { 
        fprintf(fp, "(%s.equals(\"%s\")) ", 
        $1, $3); }
		| SD_PRT EQ NUMBER{ 
        fprintf(fp, "(%s == %d) ", $1, $3); };
\end{lstlisting}

Each string in the input XML file that is not matched to the designed tokens and rules will cause an error to the program and termination of the program with an error message.

\section{Experiments}\label{sec:experiments}
When our LEX and YACC input files are ready, we can generate the final translator by calling the following commands. We have used equivalent versions of LEX and YACC called FLEX and bison respectively. In the following command xmlparser.l and xmlparser.y are the lexer and parser grammar respectively:

\texttt{flex xmlparser.l}

\texttt{bison -dy xmlparser.y}

\texttt{gcc lex.yy.c y.tab.c -o xmlparser.exe}

The translator file is named xmlparser.exe. In order to generate the Java source code for a desired XML file we can use the following command in a console command line:
\texttt{xmlparser.exe Demo.xml}

The output of this command is a file named Demo.java containing a Java class for the control policies of the network. This file can be pushed to the controller machine and compiled to be used for network management.

We used Mininet  which is a network emulator to examine the behavior of the proposed translator. Mininet supports OpenFlow, so we can register the generated Java source code with a Floodlight controller and attach the new controller to a mininet topology. Here we explain how to setup the experiments: After generating Java source code form XML file, it should be added to the Floodlight controller. We use Floodlight source code and add the generated class to its main package. It is needed to set Floodlight to load this class at startup. To do that, we first tell the loader that this module exists by adding the fully qualified module name on a configuration file named net.floodlight.core.module.IFloodlightModule in the path \texttt{src/main/resources/META-INF/services}. Then we tell the module to be loaded. So we need to modify the Floodlight module configuration file to append the Demo file. The full name of this class should be added to the file src/main/resources/floodlightdefault.properties. Now we can compile the Floodlight project and have an OpenFlow controller with our designed services running over it.

The next step is running Mininet. We setup a simple network topology containing one switch (s0) and three hosts (h1, h2, h3) connected to the switch. We also set the controller to be remote. This configuration can be obtained by this command: 

\texttt{sudo mn --controller remote --topo single,3}. 

We can test our module by calling ping command. Since our first rule was forwarding those packets going to 10.0.0.2 to port 1, so a ping from h3 to h2 or h1 to h2 will install that rule on switches and we will not receive reply from h2 while other hosts can ping each other like p1 and p3. By this simple rule we can block the traffic to h2. Running command pingall we get 33\% packet loss that means h2 is not responding to none of ping requests. 

Table \ref{tbl:results} below shows the comparison of the XML file versus generated Java file in terms of number of code lines and size of code (Kilo Byte). Using this system, we reduce coding efforts to about 1/8 of the original Java source code while the size of rule description is about 1/5 of the generated source code.

\begin{table}
\centering
\caption{Comparison of XML and Java source codes.}
\label{tbl:results}
\begin{tabular}{lcc}
\hline
                   & \multicolumn{1}{l}{\# Line of Code} & \multicolumn{1}{l}{Size (Kilo Byte)} \\ \hline
\textbf{XML file}  & 18                                  & 1                            \\ \hline
\textbf{Java file} & 147                                 & 5                            \\ \hline
\end{tabular}
\end{table}

\section{Conclusion}\label{sec:conclusion}
This paper describes a new approach of software-defined networks and presents a higher level of abstraction compared to current software-defined network programming languages. We defined a scripting language based on XML notation by which network administrators can define control policies without concerning the complexities of underlying controller framework. Indeed, this will make software-defined networking easier and more attractive for network administrators.

This work opens up the opportunity for using service oriented architecture and web services as a model of collaboration between SDN controllers (e.g., a controller offers load balancing or firewalling). 

Extending this work to support more APIs and more complicated scenarios is intended to be done in the future works. Technically, examining and manipulating other OpenFlow headers is possible by this approach. Supporting more controller frameworks with different languages is also another direction for future works.

\bibliographystyle{IEEEtran}

\bibliography{references}

\end{document}